\documentclass[aps,prl,twocolumn,reprint,showpacs]{revtex4}

\usepackage[dvips]{graphicx}
\usepackage[export]{adjustbox}
\usepackage{subfigure}
\usepackage{amsmath}
\usepackage{wasysym} 
\usepackage{MnSymbol}
\usepackage{mathtools} 
\usepackage[usenames,dvipsnames]{color}

\begin{document}

\title{On the Relaxation of Turbulence at High Reynolds Numbers}   

\author{Michael Sinhuber}

\author{Eberhard Bodenschatz}

\author{Gregory P. Bewley}
\affiliation{Max Planck Institute for Dynamics and Self-Organization, 37077 G\"{o}ttingen, Germany}


\date{\today}

\begin{abstract}

Turbulent motions in a fluid relax 
at a certain rate once stirring has stopped.  
The role of the most basic parameter in fluid mechanics, the Reynolds number, 
in setting the relaxation rate is not generally known.  
This paper concerns the high-Reynolds-number limit of the process.  
In a 
classical grid-turbulence 
wind-tunnel experiment that 
both reached higher Reynolds numbers than ever before and covered a wide range of them 
($10^4 < Re = UM/\nu < 5\times10^6$), 
we measured the relaxation rate 
with the unprecedented precision of about 2\%.  
Here $U$ is the mean speed of the flow, $M$ the forcing scale, 
and $\nu$ the kinematic viscosity of the fluid.  
We observed that the relaxation rate was Reynolds-number independent, 
which contradicts some models and supports others.  

\end{abstract}

\pacs{47.27.Gs, 47.27.Jv}

\maketitle


Turbulence dissipates kinetic energy, 
and the easiest way to see this is to let turbulence 
relax, or freely decay, 
by removing the agitation that initially set the fluid in motion.  
In so doing, 
one observes qualitatively that the fluid comes to rest.  
The rate at which that happens is the subject at hand, 
and underlies general turbulence phenomena and modeling.  
To simplify the problem, the statistics of the turbulent fluctuations 
are often arranged to be spatially homogeneous and isotropic 
\cite{karman:1938,kolmogorov:1941}.  
Even under these conditions, it is not possible quantitatively to predict the decay rate; 
it is even difficult precisely to measure the decay rate 
\cite{mohamed:1990,skrbek:2000,hurst:2007,krogstad:2010}.  
Yet empirical constraints on the properties of decaying turbulence 
are what is needed to advance the knowledge of the subject.  

The physics that control the decay are not fully understood, 
to the extent that the role of the most basic parameter in fluid mechanics, 
the Reynolds number, is not clear.  
What is known is that toward very low Reynolds numbers, 
the decay rate increases \cite{ling:1970,perot:2006}, 
as it also does when the spatial structure of the turbulence 
grows to the size of its container \cite{stalp:1999,skrbek:2000}.  
Here, we study the rate of decay in the limit of large Reynolds numbers 
for turbulence free from boundary effects,  
a rate which cannot be determined from the previous data.  


One theoretical framework predicts that the decay rate depends 
on the large-scale structure of the flow, and not on the Reynolds number 
once turbulence is fully developed 
\cite{kolmogorov:1941,saffman:1967,eyink:2000,davidson:2011,meldi:2011}.  
As the Reynolds numbers diverge to infinity, 
the scale at which energy is dissipated grows arbitrarily small, 
but these scales continue to dissipate energy as quickly 
as it is transferred to them by the large scales.  
This picture is compatible with 
an emerging consensus that the initial structure of the turbulence sets its decay rate, 
even when the flow is homogeneous 
\cite{george:1992,lavoie:2007,valente:2012,thormann:2014}.  

There is also, however, a line of thinking that an elegant and fully self-similar decay 
emerges in the limit of high Reynolds number 
\cite{dryden:1943,lin:1948,hinze:1975,george:1992,speziale:1992,burattini:2006,lavoie:2007,kurian:2009}.  
In this description, turbulence 
tends to become statistically similar to itself, when appropriately rescaled, even as it decays.  
Observation of a tendency toward slower decay with increasing Reynolds numbers 
would support this view 
\cite{george:1992,speziale:1992,burattini:2006,kurian:2009}.  
Our data, however, do not show this tendency.  

The issue is of practical consequence 
because decaying turbulence is a benchmark for turbulence models 
\cite{george:2001,hughes:2001,kang:2003,perot:2006}.  
Not only this, but both simulations and experiments 
can be performed easily at low Reynolds numbers, 
say $Re < 10^5$; 
how well can their findings describe the high-Reynolds number flows that exist in nature, 
where often $Re > 10^6$?  
Our experiments bridge this gap, 
and also contribute to a basic understanding of the nature 
of turbulence.  
The objectives are similar to those in other recent programs in fluid mechanics, 
where asymptotic scaling behavior was sought in other types of flows \cite{he:2012,huisman:2012}.  

According to dimensional reasoning, 
it is useful to think of physical laws in terms of dimensionless numbers \cite{buckingham:1914}.  
In turbulence, these numbers include the Reynolds number, $Re$, 
and a family of others that describe the initial and boundary conditions (BCs) of the flow.  
Some set of these numbers might control the decay of the turbulence.  
To isolate $Re$ effects in our experiment, 
we held fixed the other numbers.  
That is, we fixed the BCs 
so that the large-scale structure of the flow was approximately fixed.  
We then changed $Re$ by varying the viscosity of the fluid \cite{bodenschatz:2014}.  
The ability to do this was almost unique among turbulence decay experiments, 
and made it possible 
both to cover a wider range and to reach higher $Re$ than ever before.  
In previous experiments, 
changes to the BCs and to $Re$ 
were made together, 
which conflated $Re$ effects 
with those arising from changes in the large-scale structure of the flow.


Our experiment was based on 
a tradition established by early pioneers 
of using wind tunnels with grids in them 
as instruments to discover empirically how turbulence decays 
\cite{simmons:1934,dryden:1936,batchelor:1948}.  
Grids placed at one end of the tunnel stir up the flow as it passes through them, 
so that grid turbulence can be thought of as the canonical wake flow.  
Grids with different geometries produce turbulence with different structure, 
and recent work has focused on the turbulence 
downstream of grids with novel geometries 
\cite{hurst:2007,valente:2011,krogstad:2012,thormann:2014}.  
We used a single classical grid to isolate the effects of $Re$ from those of the geometry, 
and to compare our results with those from 80 years of turbulence research in wind tunnels.  

We performed the experiments in 
the Variable Density Turbulence Tunnel (the VDTT) \cite{bodenschatz:2014}.  
The VDTT circulated both air and pressurized sulfur-hexafluoride.  
The Reynolds number was adjusted by changing the pressure of the gas, 
which changes its kinematic viscosity.  
Turbulence was produced at the upstream end of the 8.8\,m 
long upper test section 
by a classical grid, that is, 
by a bi-planar grid of crossed bars with square cross section.  
The mesh spacing, $M$, of the grid was 0.18\,m, 
and the projected area of the grid was 40\% of the cross section of the tunnel.  
A linear traverse positioned probes at 50 logarithmically spaced distances, $x$, 
between 1.5\,m and 8.3\,m downstream of the grid.  
A Galilean transformation converts the distances from the grid 
into the time over which the turbulence decayed, 
so that $t = x/U$.  
Here $U$ is the mean speed of the flow down the tunnel, 
which was about 4.2\,m/s for most experiments.  

We used hot-wire probes to obtain long traces of the component of the velocity 
aligned with the mean flow.  
At each of 50 distances from the grid 
we acquired 5\,minutes 
of data.  
The experiment was repeated at 36 Reynolds numbers 
and with three probes, 
for a total of about 
$10^8$ integral lengths of data.  
We used classical hot wires produced by Dantec Dynamics 
from both 1.25\,mm lengths of 5\,micron wire, dubbed the P11 probe, 
and from 450\,micron lengths of 2.5\,micron diameter wire, dubbed the Mini probe, 
as well as the new NSTAP probe developed at Princeton that is just 60\,microns long 
\cite{bailey:2010,vallikivi:2011}.  
The probes were approximately at the centerline of the tunnel, 
with the P11 downstream of a grid bar, the Mini behind the edge of a grid bar, 
and the NSTAP between grid bars.  
The small differences between the results given by different probes 
do not affect the conclusions of this paper, and will be the subject of future detailed report.

\begin{figure}
\subfigure{
\label{fig:decaypart}
\includegraphics[width=3in,center]{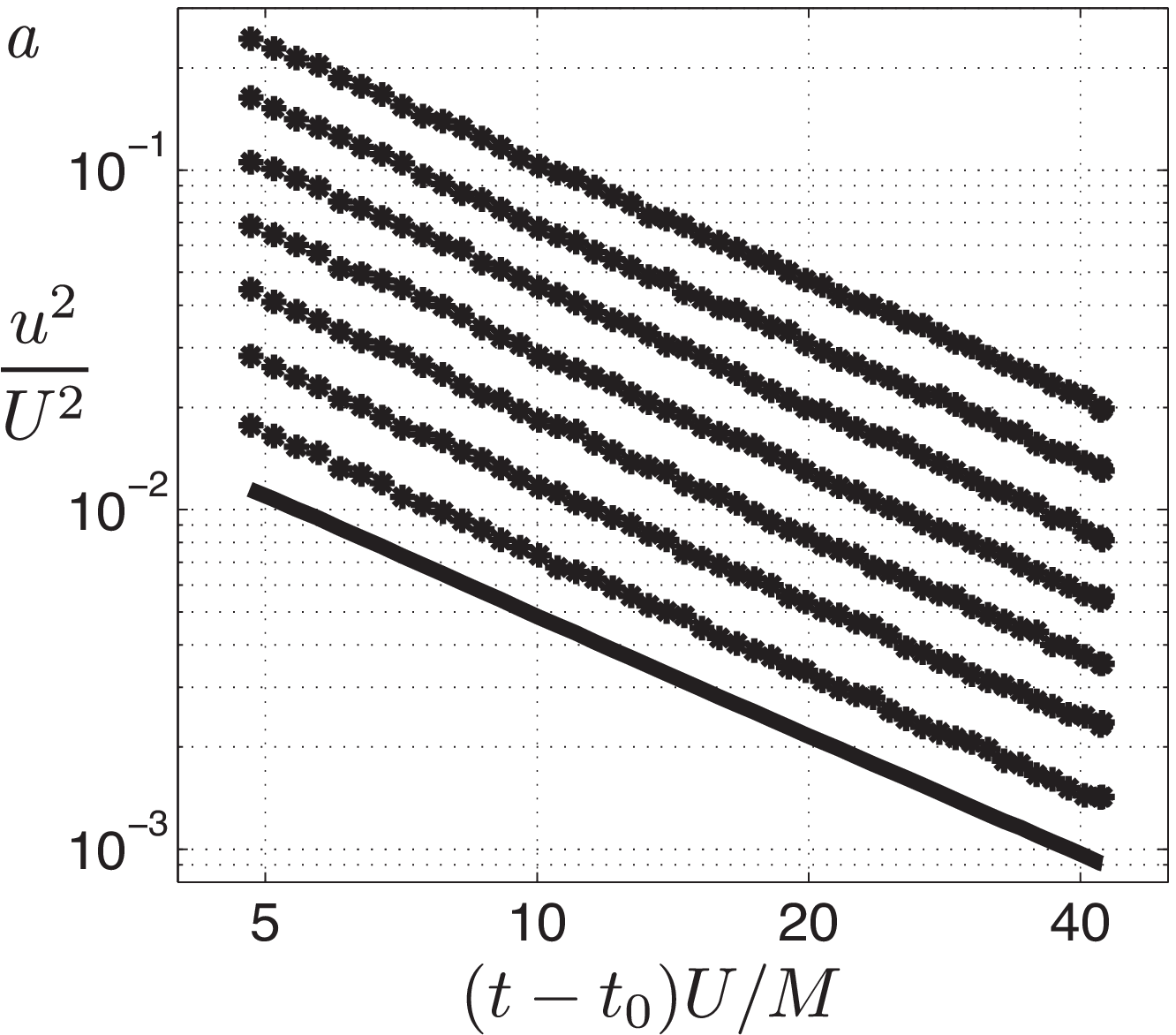}
}
\subfigure{
\label{fig:virtualoriginpart}
\includegraphics[width=3in,center]{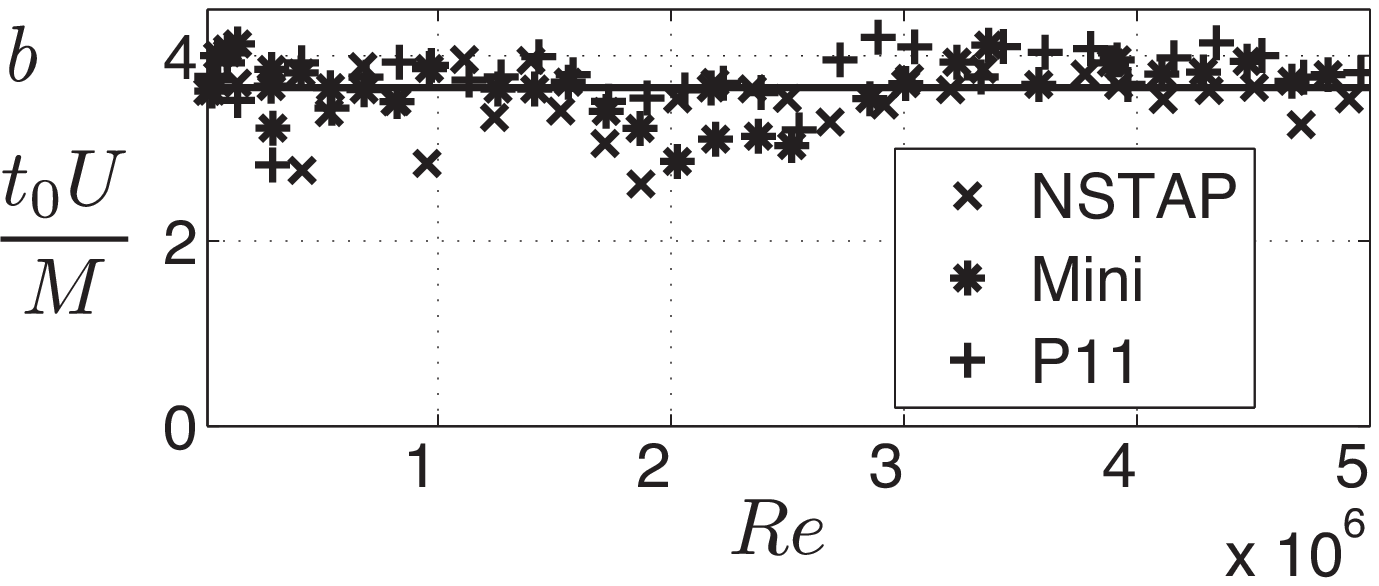}
}
\caption{(a) 
The decay of the turbulent kinetic energy, $u^2$, 
for 7 representative Reynolds numbers as a function of time.  
The linear relationship between the data on the double-logarithmic scales of the figure 
indicates a power law dependence on time.  
The values of $Re \times 10^{-3}$ for each set of data were 
26, 54, 140, 410, 820, 1700, 3200, and 4800 
from bottom to top.  
Each set of data is shifted upward 
to separate it from the lowest curve, which is unshifted.  
The lowest curve, the master curve, is 
the mean of 99 sets of data acquired at different Reynolds numbers.  
We normalized the kinetic energy by that in the mean flow, $U^2$, 
and time by $M/U$.  
Time was offset by $t_0$, the virtual origin.  
(b) We found the virtual origin 
by fitting each decay curve to a power law as described in the text.  
Over the full range of Reynolds numbers, the virtual origin fluctuated 
with a standard deviation of less than 10\% of its mean value of 
$t_0U/M = 3.66$.  
}
\label{fig:decay}
\end{figure}


Figure~\ref{fig:decay}a shows 
for several Reynolds numbers 
the normalized turbulent kinetic energy 
as a function of the time since the flow passed through the grid.  
The variance of the velocity, $u^2$, 
is proportional to the total kinetic energy in the turbulent fluctuations, 
and we refer in this paper to $u^2$ as the kinetic energy itself.  
This is because the total mass of the fluid was fixed, 
because grid turbulence is nearly isotropic, 
and because the residual anisotropy 
decays much more slowly than the energy 
\cite{lavoie:2007,kurian:2009,krogstad:2010,krogstad:2011,thormann:2014}.  
The solid curve, the master curve, 
is the mean of 99 decay curves 
accumulated by all three probes at all Reynolds numbers.  
The decay curves are remarkably similar to each other, 
even though their Reynolds numbers 
span more than two orders of magnitude.  

We see in Fig.\;\ref{fig:decay}a 
that the data follow straight lines, which is to say that they obey power laws of time, 
\begin{equation}
\frac{u^2}{U^2} = C\left((t-t_{0})\frac{U}{M}\right)^{n}, 
\label{eq:powerlaw}
\end{equation}
where $n$ quantifies the decay rate that is the subject of this paper.  
The time origin, $t_0$, for the power law, 
called the virtual origin, is not directly measurable and needs to be determined by some means.  
One way to do this is 
by 
a three-parameter fit of the data to Eq.~\ref{eq:powerlaw} 
using a nonlinear least-squares algorithm \cite{more:1978}.  
However, the uncertainties in $t_0$ and $n$ are coupled, 
leading to unreliable estimates of these parameters 
\cite{mohamed:1990,skrbek:2000,hurst:2007,krogstad:2010}.  
We describe two ways to avoid this difficulty below, 
though the conclusions of the paper are not sensitive to the method of analysis.  


\begin{figure*}
\includegraphics[width=6.9in]{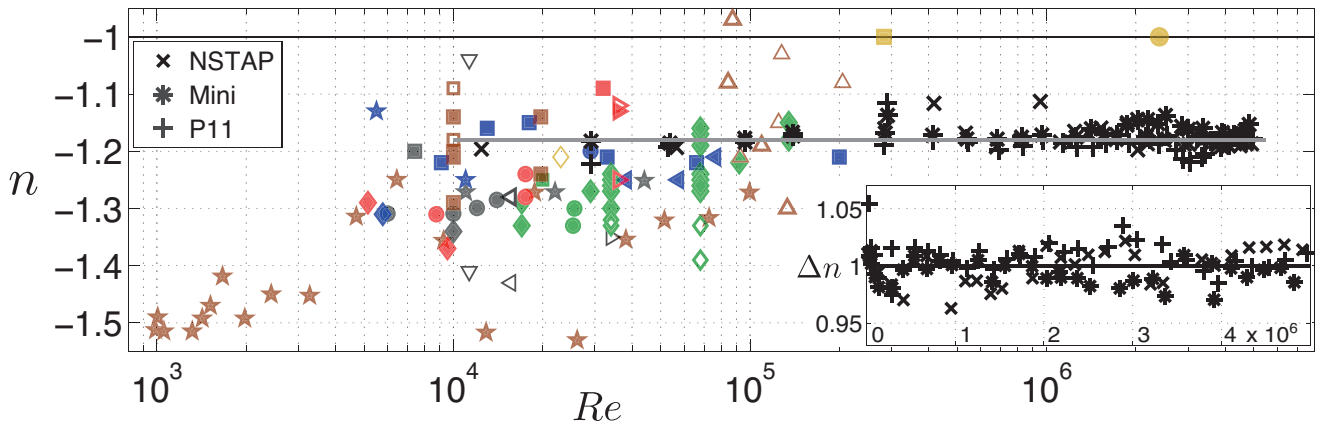}
\caption{The exponents, $n$ in Eq.\;\ref{eq:powerlaw}, 
of power-law fits to the data 
quantify the energy decay rate.  
The line is a fit to our data, 
({\large\boldmath$\times$}, {\large\boldmath$\mathrlap{\times}+$}, {\large\boldmath$+$}), 
for which the mean decay rate was 
$n = -1.18$ with a standard deviation of $0.02$.  
The 95\% confidence interval for each value of $n$ ($\pm0.9$\%)
is given 
approximately by the size of the symbols.  
Various additional data were drawn from the literature as follows: 
{\large\color{Brown}$\filledstar$} \cite{kurian:2009}, 
{\large\color{blue}$\filledstar$} \cite{batchelor:1948}, 
{\large$\filledstar$} \cite{wyatt:1955}, 
{\large\color{red}$\filleddiamond$} \cite{sirivat:1983}, 
{\large\color{blue}$\filleddiamond$} \cite{yoon:1990}, 
{\large$\filleddiamond$} \cite{warhaft:1978}, 
{\large\color{ForestGreen}$\filleddiamond$} \cite{comte:1966}, 
{\large\color{Goldenrod}$\filleddiamond$} \cite{mydlarski:1996}, 
$\filledmedsquare$ \cite{sreeni:1980}, 
{\color{blue}$\filledmedsquare$} \cite{white:2002}, 
{\color{Brown}$\filledmedsquare$} \cite{lavoie:2007}, 
{\color{ForestGreen}$\filledmedsquare$} \cite{antonia:2003}, 
{\color{red}$\filledmedsquare$} \cite{vanDoorn:1999}, 
{\color{Goldenrod}$\filledmedsquare$} \cite{bewley:2007}, 
{\footnotesize$\newmoon$} \cite{mohamed:1990}, 
{\footnotesize\color{red}$\newmoon$} \cite{uberoi:1966}, 
{\footnotesize\color{ForestGreen}$\newmoon$} \cite{vanatta:1968}, 
{\footnotesize\color{blue}$\newmoon$} \cite{uberoi:1963}, 
{\footnotesize\color{Goldenrod}$\newmoon$} \cite{kistler:1966}, 
$\filledmedtriangledown$ \cite{poorte:2002}, 
$\filledmedtriangleleft$ \cite{makita:1991}, 
{\color{red}$\filledmedtriangleright$} \cite{krogstad:2011}, 
{\color{blue}$\filledmedtriangleleft$} \cite{valente:2011}, 
{\color{Brown}$\filledmedtriangleup$} \cite{thormann:2014}. 
Closed symbols mark experiments that employed classical grids, as ours did, 
whereas open symbols mark those that employed modified grids such as fractal grids.  
Symbols with thin lines mark active-grid experiments.  
In some of these experiments, 
modifications to the grids 
were made deliberately 
to elicit changes in the decay \cite{lavoie:2007,thormann:2014}.  
The inset shows on a linear scale of $Re$ 
the relative decay exponents 
as described in the text.  
Our data reveal that the rate of decay is invariant with respect to the Reynolds number 
for $Re \gg 10^4$.  
}
\label{fig:decayexponent}
\end{figure*}

Figure~\ref{fig:decay}b shows 
that the virtual origin, to a first approximation, was $Re$ independent.  
Its value is probably connected to the development of the wakes 
behind the bars that compose the grid; 
in previous experiments different grid geometries produced different virtual origins 
\cite{comte:1966,lavoie:2007,thormann:2014}. 
The wakes behind square bars 
are nearly $Re$ independent in the 
$Re$ regime of our experiments, 
and do not show the drag crisis 
characteristic of circular cylinders \cite{schewe:1984}.  
In view of these observations, it seems reasonable to 
assign a fixed value to the virtual origin.  

With the virtual origin fixed to its mean value, 
$t_0U/M = 3.66$, 
we could measure the decay exponents more precisely 
with a two- (rather than three-) parameter fit of the data to Eq.\;\ref{eq:powerlaw}.  
Here, $C$ and $n$ were free parameters, 
and we use the power law as a tool 
to uncover the $Re$ evolution of the decay process.  

Figure~\ref{fig:decayexponent} 
shows our measurements of the decay exponents 
and compares them to those in the literature.  
Our data 
neither approach $n = -1$ nor trend toward a slower decay 
with increasing $Re$, 
despite the wide range of high $Re$ achieved in the VDTT.  
Combining all our data, we obtain a mean decay exponent of $n = -1.18\pm0.02$, 
close to Saffman's prediction of $-1.2$ 
 \cite{saffman:1967}.  

Our second method of analysis 
emphasizes the $Re$ variation of the decay rate, 
rather than its specific form.  
We make the Ansatz that the kinetic energy decay at one Reynolds number, 
$E_i\left(t,Re_i\right)$, 
is a power law function of the decay at another Reynolds number, 
$E_j\left(t,Re_j\right)$, 
so that $E_j \sim E_i^{\Delta n}$, 
and $\Delta n$ is a \textit{relative} decay exponent.  
Such a relationship does not strictly require power-law evolution 
of the energies, $E_i$ or $E_j$ \cite{benzi:1995}.  
If, however, such a power-law decay does hold, 
as in our experiment, 
then $\Delta n = n_j/n_i$, 
where $n_i$ and $n_j$ are the decay exponents at the two Reynolds numbers.  
Note that the variation in the virtual origin must be small relative to its mean, 
as it was in our experiment.  
In practice, 
we took $E_i$ to be the master curve shown in Fig.\;\ref{fig:decay}a, 
$E_j$ to be the decay of $u^2$ at the various $Re$, 
and we extracted values of $\Delta n$ 
by fitting power laws to graphs of $E_j$ against $E_i$.  
The advantage was not only that the ambiguity of finding the virtual origin was eliminated, 
but the requirement of power-law decay was relaxed.  

The inset of Fig.\;\ref{fig:decayexponent} 
presents the relative scaling exponents derived by the above procedure.  
Here, a trend toward a slower decay would appear as a tendency of 
the data toward lower values with increasing $Re$.  
Again, the data show no such trend and scatter around 1.0, 
which means that the dominant behavior is for the decay 
not to change with increasing $Re$.


Some comments on the data 
gathered from the literature in Fig.\;\ref{fig:decayexponent} follow.  
Different conclusions may be drawn from these data: 
an approach to an $n = -1$ decay 
might be seen in the green diamonds \cite{comte:1966,george:1992}, 
while the blue squares may be consistent with no change \cite{white:2002}.  
In computer simulations \cite{burattini:2006,ishida:2006,perot:2011}, 
the decay slowed down slightly with increasing $Re$, 
but with similar reach in $Re$ and scatter in the exponents 
as in previous experiments.  
All in all, this scatter is significant, 
and may be attributable to 
variations in initial and boundary conditions between experiments, 
that is, to the geometry of the grids used or to other variations in the experimental design.  
The results from some experiments 
are not shown either because the form of the decay was not a power law 
or because the exponents fell out of the range of the plot.  
Collectively, the previous data leave open the question of how the decay rate 
behaves in the limit of large $Re$.  

Because the Reynolds number seemed not to 
govern the decay rate in our experiment, 
we sought an explanation for the decay rate in terms of the large-scale structure of the flow.  
One way to derive predictions for the decay rate is to consider the evolution 
equation for the kinetic energy in freely decaying turbulence: $(3/2)du^2/dt = -\epsilon$.  
In the classical description, the dissipation rate, $\epsilon$, 
is independent of $Re$ 
so that $\epsilon= C_\epsilon u^3 / L$, 
and $(3/2)du^2/dt = -C_\epsilon u^3 / L$.  
The constant, $C_\epsilon$ is of order one in most flows \cite{sreeni:1998} 
and it was also in our experiment, 
varying over time by about 4\%.  

In order to integrate the energy equation, 
some relationship between the energy, $u^2$, and the correlation length, $L$, 
must be derived.  
Typically, this relationship is a power law, $L \sim \left(u^2\right)^m$, 
with $m = -1/2$, $-1/3$, and $-1/5$, 
for each of the self-similar, Saffman, and Kolmogorov theories, respectively
\cite{kolmogorov:1941,dryden:1943,saffman:1967}.  
Yet other exponents can be supported by different arguments 
\cite{eyink:2000,davidson:2011}.  
Integration of the energy equation then yields predictions 
for the decay law, $u^2 \sim t^n$, with $n=-1$, $-6/5$ and $-10/7$, 
for the three theories, respectively, 
and $n = 2 / (2m - 1)$ generally \cite{krogstad:2010}.  
Fortunately, 
it is possible to look directly for a relationship between 
$u^2$ and $L$, 
without the ambiguity of determining a virtual origin, 
since the virtual origin must be a property of the flow and hence be the same for all of its statistics.  

\begin{figure}
\includegraphics[width=3in]{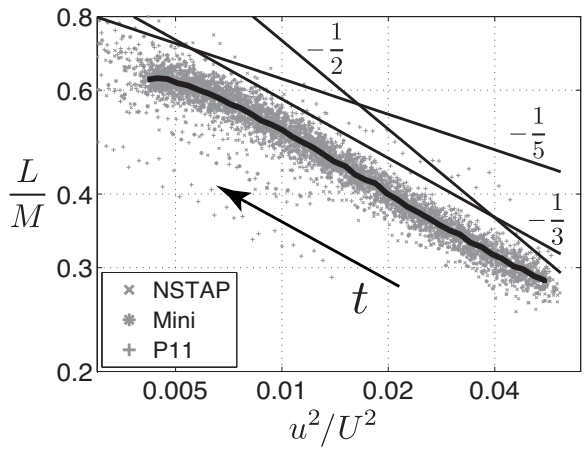}
\caption{The correlation lengths, $L$, 
were approximately a power law of 
the kinetic energy, $u^2$, during the decay.  
Time proceeds from right to left in the graph, 
as indicated by the arrow.  
Here we combined all data without discrimination 
since there was no trend with respect to $Re$.  
To produce the smooth curve, 
we took the median values of $L/M$ and $u^2/U^2$ data falling within 
logarithmically spaced bins in $u^2/U^2$.  
The relationships posited by well-known theories are indicated by straight lines.  
Our data are consistent with Saffman's theory, with slope $-1/3^{rd}$.  
}
\label{fig:lvsu}
\end{figure}

Figure~\ref{fig:lvsu} shows the correlation length, $L$, 
calculated from the scale, $r$, at which the velocity correlation function dropped below $1/e$.  
The velocity correlation function was $f(r) = \langle u(x) u(x+r) \rangle_x / u^2$, 
which could be calculated from our time series of velocity 
in the usual way through Taylor's hypothesis \cite{taylor:1938}.  
Grey symbols mark our data, whereas the solid lines denote the various theoretical predictions.  
Neither Kolmogorov's nor the self-similar decay agrees with our data, 
whereas Saffman's prediction is consistent with our data for large $u^2/U^2$.  

Where the data are approximately straight 
in Fig.~\ref{fig:lvsu}, 
a power-law fit yields $-0.35$ 
so that the theory predicts $n = 2 / (2m - 1) = -1.18$, 
which is 
precisely what we measured directly from the decay of kinetic energy.  
We attribute the deviations from Saffman's prediction at large times 
to confinement of the flow by the walls 
of the tunnel \cite{stalp:1999,skrbek:2000}.  

We now give a brief physical interpretation of Saffman's theory.  
We first picture turbulence 
as a sea of vortices 
that each carry both linear and angular impulses \cite{davidson:2004}.  
The vortices interact as the turbulence evolves, and while their kinetic energies decay, 
their linear and angular impulses are conserved.  
The conservation of their linear impulses 
is embodied in Saffman's invariant \cite{birkhoff:1954,saffman:1966}.  
This invariant 
yields the relationship between $u^2$ and $L$ mentioned above, 
and so to the $-6/5^{ths}$ decay law \cite{saffman:1967}.  
In the case that the linear impulses of the vortices are 
initially small 
compared with their angular impulses, 
the character of the decay is dominated by the conservation of the latter, 
which is embodied in Loitsyanskii's invariant \cite{loitsyanskii:1939} 
and Kolmogorov's faster $-10/7^{ths}$ decay law \cite{kolmogorov:1941}.  
In other words, turbulence is more durable when its constituent 
vortices carry significant linear impulse 
in addition to their angular impulse.  

In summary, 
we observed the decay rate of turbulence to be Reynolds number independent, 
and for the decay to proceed in a way consistent with Saffman's predictions \cite{saffman:1967}.  
The data do not agree with the predictions of 
Kolmogorov, Dryden, George, Speziale, Bernard 
and others \cite{kolmogorov:1941,dryden:1943,george:1992,speziale:1992}.  
Initial and boundary conditions 
different from those in our experiment, 
established for instance by active or fractal grids \cite{mydlarski:1996,valente:2011},  
might produce different decay rates 
and even different $Re$ dependencies.  
Our results, however, indicate that the self-similar decay is probably not 
the generic high-$Re$ decay.  
We attribute the residual $Re$ dependencies in our data 
to minor $Re$ variations in the 
turbulence production by the grid, 
and not to an effect of $Re$ on the physics of the decay.  
In closing, 
we note that a deeper understanding of the decay mechanisms 
may be acquired by looking at different systems.  
For example, 
quantum turbulence, 
whose small-scale physics are completely different, 
may have the same large-scale structure as classical turbulence 
and may relax in the same way \cite{stalp:1999,skrbek:2012}.  


\begin{acknowledgments}
The VDTT would not have run without 
A. Renner, A. Kopp, A. Kubitzek, H. Nobach, U. Schminke, 
and the machinists at the MPI-DS 
who helped to build and maintain it.  
The NSTAPs were designed, built and graciously provided by 
M. Vallikivi, M. Hultmark and A. J. Smits at Princeton University.  
We are grateful to F. K\"ohler and L. Hillmann for help acquiring the data 
and to G. Schewe 
and our many inspiring colleagues 
for discussions.  
\end{acknowledgments}

\bibliographystyle{apsrev}

\end{document}